\documentclass[11pt]{article}

\usepackage[margin=1in]{geometry}
\usepackage{amsmath,amssymb}
\usepackage{booktabs}
\usepackage{array}
\usepackage{tabularx}
\usepackage{float}
\usepackage{needspace}
\usepackage{graphicx}
\usepackage{xcolor}
\usepackage{microtype}
\usepackage[numbers,sort&compress]{natbib}
\usepackage[hidelinks]{hyperref}
\usepackage{xurl}
\usepackage{enumitem}

\newcolumntype{Y}{>{\raggedright\arraybackslash}X}
\newcommand{\methodname}{Provider Behavior Profile}
\newcommand{\tv}{\operatorname{TV}}
\newcommand{\ppmi}{\operatorname{PPMI}}
\newcommand{\vect}[1]{\boldsymbol{#1}}
\definecolor{caseblue}{HTML}{315B7D}
\newenvironment{caseexample}[1]{%
  \Needspace{7\baselineskip}\par\medskip\noindent
  \begin{minipage}{\linewidth}
  \color{caseblue}\rule{\linewidth}{0.9pt}\par\vspace{0.35em}
  \textbf{\MakeUppercase{#1}}\par\smallskip\color{black}%
}{%
  \par\vspace{0.35em}\color{caseblue}\rule{\linewidth}{0.45pt}
  \end{minipage}\par\medskip
}

\title{Explainable Provider Behavior Profiles\\
for Fraud, Waste, and Abuse Review}
\author{Yubin Park, PhD\\
\textit{Falcon Health, Inc.}
\and
Evan Brociner\\
\textit{Falcon Health, Inc.}}
\date{September 2026}

\begin{document}

\maketitle

\begin{abstract}
Claims can show that provider behavior changed, but claims alone cannot establish why it
changed. Fraud, waste, and abuse (FWA) review therefore requires investigation: analysts
must identify material behavior, locate the codes and dollars that drive it, and test
plausible clinical, operational, and administrative explanations. The objective is not to
predict future misconduct or infer intent. It is to maintain a reproducible description of
provider scale, service composition, temporal change, and observation context so that an
analyst can see why a provider entered a review queue. Earlier development experiments
considered forecast residuals as one possible signal, but those residuals combined growth,
service-line changes, code maintenance, payer selection, incomplete observation, and
potentially concerning behavior. We therefore evaluate the operational representation by
coverage, retrieval utility, refresh stability, cross-context replication, and explanation.

We instead formulate provider review as a descriptive representation problem. For
provider $i$, procedure $j$, and period $t$, we write observed amount as
$y_{ijt}=s_{it}p_{ijt}$, where $s_{it}$ measures provider scale and $p_{ijt}$ describes
procedure composition. The profile records scale history, effective-dated code lineage,
clinical-family shares, attributed mix change, first-use events, billing context, and the
difference between broad Medicare and client-specific views. An optional rank-32
nonnegative factorization of positive procedure co-occurrence supplies a fixed semantic
geometry for similarity and retrieval. The representation surfaces evidence for review;
it does not infer intent or adjudicate FWA.

This profile is one engine within Falcon's broader review system, which combines it with
complementary rules, peer and network context, statistical models, and clinical or coding
review; its outputs are designed to feed and extend those other engines. We describe it
independently here because its measurement contract and explanations can be evaluated on
their own merits.

The refreshed ten-quarter proprietary Medicare build covers Carrier and DME and contains
1.27 million providers and 9,212,786 provider-quarter profiles through 2026 Q2. The selected rank-32
semantic dictionary represents 3,641 procedures and improves overall held-out recall at
10 from 36.0\% under popularity ranking to 44.4\%; for 230 high-cost rare events it reaches
23.5\% recall at 10 and 56.1\% at 50, versus zero for popularity. Under the governed
materiality and service-breadth eligibility contract, 414,093 providers enter the current
national review population. Adjacent-quarter review-rank correlations are 0.262 and 0.237,
while 81.8\% and 90.8\% of the prior eligible populations remain eligible. A common
completed-quarter replication across eight client
panels produces 894,849 profiles and preserves amount-weighted semantic coverage of
86.9--95.2\%. The results support a layered architecture in which transparent descriptions
form the core and learned representations add optional semantic context. A separately
frozen retrospective comparison with administrative endpoints remains a secondary
construct check; it is neither the objective of the algorithm nor evidence of fraud.
\end{abstract}

\section{Introduction}

Labeling a provider as fraudulent requires rigorous investigation and, where applicable,
formal administrative or legal process. Claims can reveal billing behavior, but claims
alone rarely establish intent; treating a statistical pattern as intent would be
speculation. The practical analytical task is therefore to identify which providers merit
review and what evidence an investigator should examine next. Confirmed fraud labels
remain rare, delayed, and shaped by enforcement capacity. Administrative lists observe
only a process- and capacity-selected fraction of potentially concerning conduct, while
some listed actions concern non-fraud compliance failures. They are therefore neither
exhaustive fraud labels nor clean ground truth. Claims also reflect clinical need, coding rules, reimbursement
policy, organizational relationships, payer selection, and data-processing conventions.
The same statistical change can therefore represent intentional overbilling, a new service
line, employment reassignment, an HCPCS replacement, incomplete reporting, or a payer
observing only one part of a diversified practice. CMS similarly distinguishes improper
payments from fraud and warns against reading payment-error estimates as fraud rates
\citep{cms_improper_payments_2025}, and GAO continues to list Medicare among the federal
programs most vulnerable to improper payment and fraud risk \citep{gao2025highrisk}.

Prediction offers one natural starting point. A model estimates expected future
utilization, and a large residual identifies a possible review candidate. We tested that
approach with persistence, autoregressive, specialty-pooled, hurdle, and low-rank models.
The immediately preceding observation contained most of the estimable signal. More model
structure produced limited or unstable improvement, especially for code entry and exit.
The residual also answered the wrong question: it measured forecast error without telling
the analyst whether growth, substitution, code-system maintenance, client selection, or a
potentially concerning change caused it. Models trained on broad Medicare behavior did not
transfer cleanly to client claims because each client observes a selected practice segment,
not a random sample of the same process.

Prior research places these findings in context. Unsupervised FWA studies use residuals,
provider profiles, peer comparisons, latent procedure groups, graphs, and domain rules to
rank cases without complete fraud labels \citep{musal2010unsupervised,ekin2019bayesian,
shekhar2023explainable}. Their most useful outputs often describe the exact codes, peer
differences, and dollars behind a flag. Simple domain-informed measures can also produce
credible review signals; for example, implied service time identifies billing volumes that
deserve investigation without predicting fraud \citep{fang2017hours}. These studies suggest
that prediction and anomaly detection work best as components of a broader evidence
system.

The method developed here is likewise built to run as one engine within Falcon's broader
FWA review system. Candidate identification, funneling, and filtering combine this
profile with complementary rules, peer and network context, statistical models, and
clinical or coding review, and the profile's outputs are designed to feed those other
engines. This paper isolates the provider-behavior profile so that its assumptions,
empirical support, and explanations can be shared and scrutinized on their own.

We therefore treat representation, rather than point prediction, as the organizing
objective. The profile must answer five questions:

\begin{enumerate}[leftmargin=*]
  \item How large is the provider's observed practice, and how has that scale changed?
  \item Which procedures or clinical service families characterize the provider?
  \item Which components account for a change in dollars or service mix?
  \item Is a first-observed code established, related, substituted, globally new, or
        simply new to this observation system?
  \item How does client-observed behavior differ from the provider's broader Medicare
        behavior, conditional on coverage and timing?
\end{enumerate}

The resulting \methodname{} is a compact historical description. It supports analyst
review, provider retrieval, cohort construction, monitoring, peer comparison, and later
rules or rankings. It also keeps the underlying evidence visible. Throughout this paper,
``unusual,'' ``divergent,'' and ``reviewable'' describe observed patterns; they do not
allege fraud or clinical inappropriateness.

\subsection{Alternatives considered}

We evaluated five alternatives. Point-forecast residuals mixed scale and composition and
did not explain the source of change. Supervised fraud classification targeted enforcement
labels rather than provider behavior and inherited their selection and delay. Peer-outlier
scores supplied useful context but did not describe a provider's own longitudinal change.
Direct masked NMF of sparse provider shares produced higher held-out error than a code-mean
baseline and unstable procedure vectors. Provider-specific transition matrices required
more history than eight persistent quarters could supply. These results support a layered
design: transparent behavior first, authoritative lineage second, peer and cross-view
context third, and learned semantic geometry where it adds demonstrated value.

\begin{figure}[ht]
  \centering
  \includegraphics[width=\linewidth]{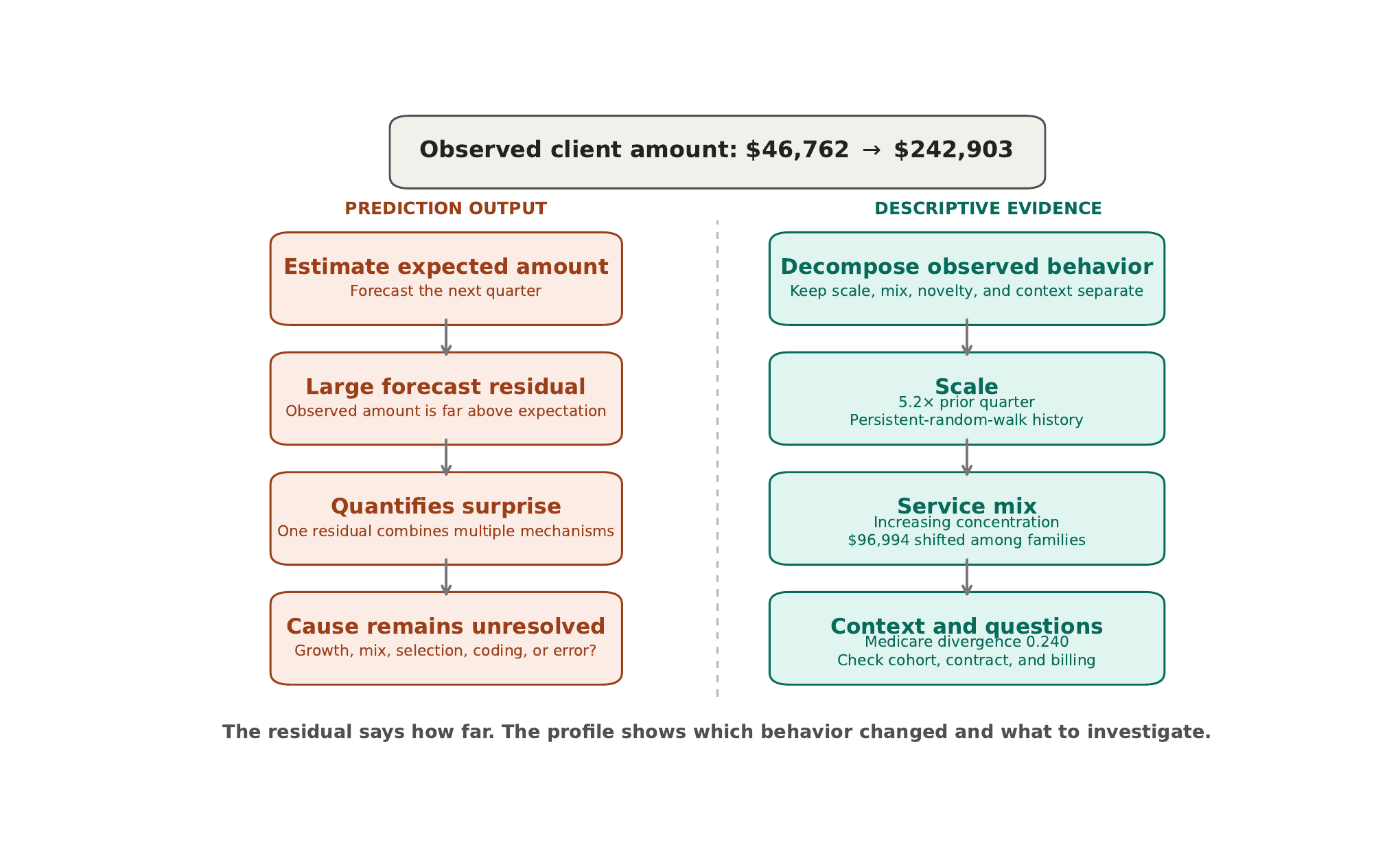}
  \caption{A forecast residual identifies surprise but does not identify its source. A
  behavior profile separates scale, service-mix, and cross-view evidence while retaining
  the amounts and classes needed to formulate review questions. Values shown are an
  illustrative rendering of a supported Carrier+DME case.}
  \label{fig:prediction-description}
\end{figure}

Figure~\ref{fig:prediction-description} makes the central design choice concrete. A
prediction path quantifies how far an observed amount departs from an estimate, whereas
the descriptive path separates which dimensions changed and which follow-up questions
remain.

\subsection{Contributions}

This work contributes: (i) a scale--composition decomposition tailored to investigation;
(ii) an effective-dated, lineage-aware family representation with attributed temporal
change; (iii) an optional fixed procedure geometry learned from positive co-occurrence
rather than censored absences; (iv) a shared coordinate system for broad Medicare and
client-specific views without transferring a forecast equation; and (v) an empirical
simplicity test that promotes the smallest representation that preserves explanatory
value; and (vi) a rolling-origin, post-hoc comparison with two external administrative
endpoints that are never used to fit the representation or construct its scores.

\section{Background and Related Work}

\subsection{Detection, prediction, and investigation}

Healthcare FWA analytics includes rules, supervised classification, regression,
clustering, peer comparison, tree-based isolation \citep{liu2008isolation}, graphs, and
deep generative anomaly detection \citep{naidoo2020gan}. These papers share an application
domain but do not all address the same problem. Reviews distinguish supervised methods
trained on known cases from unsupervised methods that identify claims or providers for
later assessment \citep{joudaki2014review}. Early unsupervised Medicare work similarly
treated statistical output as a prescreening signal rather than proof of fraud
\citep{musal2010unsupervised}. We adopt that investigative boundary directly.

Supervised Medicare studies provide an important contrast. They train classifiers using
provider-exclusion labels \citep{bauder2017medicare}, predict a physician's specialty from
the procedures reported and treat mismatch as a detection feature
\citep{herland2020provider}, or study resampling, cost-sensitive objectives, and threshold
selection under extreme class imbalance \citep{johnson2019neural}. This literature shows
what can be optimized when a usable target label and decision unit are declared. Our
upstream task is different: construct a stable, reusable account of provider behavior
when labels are incomplete, selected, delayed, or do not identify which observed service
was problematic. We therefore cite these studies as neighboring detection work, not as
the methodological template for the profile.

Two explainable unsupervised systems are more directly connected to our approach.
Shekhar, Leder-Luis, and Akoglu combine patient-history expenditure regression, ICD
subspace outliers, and peer DRG comparisons in a multi-view hospital-ranking system
\citep{shekhar2023explainable}. Their stronger multi-view results and code-, peer-, and
dollar-level case explanations support two of our design choices: a forecast residual is
useful as one view rather than the whole provider description, and a ranking becomes
operationally meaningful only when its drivers remain inspectable. Their DOJ corpus is a
retrospective positive-unlabeled evaluation, whereas our present validation concerns the
stability and explanatory content of the representation itself.

De Meulemeester et al. provide the closest operational analogue
\citep{demeulemeester2025explainable}. They aggregate practitioner resource-use records,
learn compact representations for high-cardinality categorical fields from co-occurrence,
rank practitioners with unsupervised anomaly detectors, and combine SHAP with original
code distributions and summary statistics for expert review. This directly supports our
emphasis on provider-level representation, compact semantic context, and investigator-facing
evidence. Our profile differs by defining the explanation before the ranking: it explicitly
separates quarterly scale and service mix, maintains effective-dated code lineage, measures
first use and coverage, and compares client and proprietary Medicare views. The learned
geometry remains optional rather than serving as the anomaly model that must be explained
afterward.

\subsection{Provider representation and peer context}

Provider-profile research most directly matches our measurement objective. Ekin,
Lakomski, and Musal fit an unsupervised Bayesian hierarchy to Medicare Part B data, group
medical procedures, identify latent provider--procedure patterns, and use them for
similarity and outlier assessment \citep{ekin2019bayesian}. This is the clearest precursor
to our use of provider--procedure structure as a label-free prescreening representation.
We retain the same broad insight while adding an explicit scale--composition decomposition,
quarterly change, effective-dated lineage, and a shared Medicare/client coordinate system.

Peer comparison supplies another descriptive view. A peer distribution can show which
codes and dollars make a provider different from otherwise similar practices. It can also
confound specialization, case mix, geography, or network selection when the peer
definition is weak. We therefore keep peer distance separate from the provider's own
temporal movement and require support and coverage fields for interpretation.

Graph systems address a complementary unit of evidence. Liu et al. represent patients,
providers, pharmacies, and other entities as a heterogeneous healthcare network and find
unusual actors, relationships, temporal changes, geographies, and subgraphs
\citep{liu2015graph}. Muhammad et al. apply heterogeneous GNNs and post-hoc graph
explainers to labeled activity-level medical-claims decisions
\citep{muhammad2025gnn}. FraudAuditor instead uses patient co-visit networks, community
detection, and investigator-facing visual analysis to examine collusive health-insurance
groups \citep{zhou2023fraudauditor}. These healthcare studies support our claim that
provider behavior is only one view: relational and collusive patterns require a network
layer and may not be visible in a provider's own quarterly composition.

SEFraud is included only as broader fraud-analytics context, not as healthcare evidence.
It integrates learned node-feature and edge masks into a heterogeneous graph detector and
reports a deployed banking application \citep{li2024sefraud}. Its relevance is the design
principle that explanations can be part of the detector rather than an unrelated
afterthought. Our system applies that principle differently: the governed provider
description is independently inspectable, and future graph signals can enrich it without
replacing the longitudinal profile.

\subsection{Transparent measures and learned semantics}

Transparent domain measures provide a second direct influence. Adapted incomplete
Benford models show that a narrow behavioral
constraint can flag unusual insurance claims without fitting a general prediction model
\citep{lu2005benford}. We do not use Benford's law here; its relevance is the broader
principle that an interpretable measurement contract can itself create a review signal.
Fang and Gong offer an even closer example by translating Medicare procedure codes into
implied physician work hours and identifying approximately 2,300 physicians billing more
than 100 hours per week \citep{fang2017hours}. Their measure does not infer fraud; it maps
billing composition into a real-world constraint that an investigator can test. That
logic directly motivates our separation of observable behavior from adjudication and
future extensions such as overlapping services, units per patient, and
capacity constraints defined for individual procedures.

Finally, the representation-learning citations are methodological foundations rather
than FWA precedents. Nonnegative matrix factorization supplies an additive, parts-based
representation \citep{lee1999nmf}. PPMI supplies a count-based distributional geometry,
and careful count-based designs can compete with predictive embeddings such as skip-gram
word2vec \citep{levy2015ppmi,mikolov2013distributed}. We borrow these tools only for
procedure similarity, retrieval, sparse-vocabulary smoothing, and fixed-width machine
representation. The learned geometry does not serve as a risk score; literal codes,
clinical families, and attributed amounts preserve the explanation.

\section{Data and Study Design}

\subsection{Proprietary Medicare reference data}

The production reference panel, which we refer to as PSPS, is Falcon's proprietary
quarterly summary of Medicare Carrier and DME activity at rendering-NPI by HCPCS grain,
excluding known phantom identifiers; it is not a public CMS product or file. CMS
provider-and-service documentation is cited only to describe the general organization of
provider-service data products \citep{cms_provider_service_2026}---the observations
analyzed here are Falcon's own. The current operational audit uses ten complete quarters
from 2024 Q1 through 2026 Q2 and combines Carrier and DME under the same representation
contract. It contains 1,266,479 provider scale descriptors and 9,212,786 provider-quarter
profiles. Historical experiments that motivated individual design choices are labeled as
such; current-data claims report the governed 2026 Q2 Carrier+DME production build.

We treat absence from the proprietary summary as censored or unreported, not as a
verified zero. This matters for sparse matrix estimation: a missing provider-code pair
does not supply the same negative evidence as a fully adjudicated zero. Provider scale uses
nonnegative reported amount; we normalize procedure shares only when total
scale is positive.

\subsection{Client claims}

We use anonymized labels for eight payer panels; the labels do not encode their identities.
Each client profile uses paid
Carrier and DME professional claim lines; facility claims require a separate observation
and attribution model and are excluded. The common amount contract is paid-to-provider
amount; nonpositive line amounts do not contribute
compositional mass. Profiles use fixed calendar quarters. Completed quarters retain
observed amount. If the latest reliable service date falls before quarter end, the profile
retains quarter-to-date amount and separately records calendar-day completion and an
estimated full-quarter run rate. Each refresh rebuilds a bounded quarterly amount history
so late-arriving claims can revise prior quarters without creating competing rolling-window
definitions. The broad Medicare dictionary is read-only; client-derived state remains isolated in the
client schema and never feeds back into it, as detailed in Section~7.3.

This choice followed a production data-quality audit showing that eligibility fields can
repeat on zero-paid reporting lines and therefore should not be interpreted as provider
revenue. The representation consequently requires an explicit, versioned amount contract
rather than an interchangeable collection of source fields.

\subsection{Procedure metadata and lineage}

An effective-dated procedure concept layer combines HCPCS lifecycle fields, structured
CMS predecessor/successor crosswalks, and RBCS clinical categories. Exact authoritative
lineage takes precedence over statistical similarity. We classify a successor as an
observed replacement only when the same provider previously used the predecessor.
Clinical family and embedding proximity remain contextual and do not manufacture code
identity.

\subsection{External administrative endpoints}
\label{sec:external-endpoints}

For post-hoc external evaluation, we use the current CMS revoked-provider snapshot and
HHS OIG List of Excluded Individuals/Entities (LEIE), retaining the earliest usable
effective date per NPI. These lists observe only a selected fraction of the underlying
conduct that may warrant investigation, and their reasons include fraud-related,
licensure, reporting, enrollment, exclusion, and other administrative actions. An
unlisted provider is not a verified negative, and a listed provider is not thereby a
judicially established fraud case. No list membership, reason, date, or derived label is
used in dictionary fitting, provider representation, score construction, or queue
ranking; the lists enter only after those outputs have been frozen for evaluation.

\section{Methodology}

\subsection{Scale--composition decomposition}

Let $y_{ijt}\geq0$ be observed revenue for provider $i$, procedure $j$, and period $t$.
Define total scale and procedure share as

\begin{equation}
  s_{it}=\sum_j y_{ijt}, \qquad
  p_{ijt}=\frac{y_{ijt}}{s_{it}}, \qquad \sum_j p_{ijt}=1,
  \label{eq:decomposition}
\end{equation}

for $s_{it}>0$. Then $y_{ijt}=s_{it}p_{ijt}$. This identity separates overall growth
from redistribution among services. It also permits scale and mix to carry different
support requirements and different missing-data semantics.

\subsection{Scale descriptors}

The core scale profile stores current revenue, adjacent-period and year-over-year growth,
history length, volatility on $z_{it}=\log(1+s_{it})$, the largest historical change,
missing-period indicators, and billing-entity concentration. For diagnostic model
selection we compare:

\begin{align}
  \text{AR(0):}\quad & z_{it}=\mu_i+\varepsilon_{it},\\
  \text{random walk:}\quad & z_{it}=z_{i,t-1}+\varepsilon_{it},\\
  \text{random walk with drift:}\quad & z_{it}=z_{i,t-1}+g_i+\varepsilon_{it},\\
  \text{regularized AR(1):}\quad & z_{it}-\mu_i=\phi_i(z_{i,t-1}-\mu_i)+\varepsilon_{it}.
\end{align}

Provider AR coefficients are shrunk toward a pooled coefficient. Selection uses rolling
one-step errors and chooses the simplest model within 5\% of a provider's best log-scale
sum of squared errors. Providers without complete history receive summary descriptors,
not unsupported individual dynamics. These classes are diagnostic labels; the core
profile does not require an AR forecast.

We distinguish two outputs that are easy to conflate. The structural
\emph{scale behavior class} summarizes the provider's history: stable level, persistent
random walk, sustained growth or contraction, mean reversion, emerging-practice growth,
or low/partial history. A separate \emph{current scale signal} describes the latest
movement relative to that structure: within model range, mature-provider acceleration,
emerging-practice acceleration, or contraction/reversion. The historical class therefore
does not become a flag merely because it differs from Medicare, and the latest movement
does not redefine the provider's longer-run class.

\subsection{Lineage-aware service mix}

We map literal codes through effective-dated concepts and clinical families. Let
$g(j,t)$ denote the valid family for code $j$ at time $t$. The family composition is

\begin{equation}
  P_{ikt}=\sum_{j:g(j,t)=k}p_{ijt}.
\end{equation}

We summarize consecutive mix movement with total variation,

\begin{equation}
  \tv(P_{it},P_{i,t-1})=\frac{1}{2}\sum_k
  \left|P_{ikt}-P_{ik,t-1}\right|,
  \label{eq:tv}
\end{equation}

and by attributed dollar and share changes for the largest contributing families. Total
variation lies in $[0,1]$: zero denotes identical family shares and one denotes disjoint
support. Literal-code changes remain available so that family aggregation cannot hide a
material code substitution.

For operational interpretation, we assign a transparent mix behavior class from these
same shares. Total variation no greater than 0.10 is stable mix. A material introduced or
exited share, together with a large family gain or loss, defines an emerging service line
or service-line exit. A Herfindahl change of at least 0.15 in magnitude distinguishes
increasing concentration from increasing diversification. Total variation of at least
0.50 without a more specific entry/exit label is abrupt substitution; remaining movement
is gradual rebalancing. Providers without a comparable prior quarter receive
low/partial-history status. These labels are deterministic descriptions, not fitted
mixture-model classes.

\subsection{Optional PPMI procedure geometry}

For supported procedures, let $B_{ij}$ indicate that provider $i$ reported procedure $j$
in the reference window. Define co-occurrence $n_{jk}=\sum_i B_{ij}B_{ik}$ and marginal
$n_j=\sum_i B_{ij}$. For $N$ sampled providers,

\begin{equation}
  X_{jk}=\ppmi(j,k)=\max\left\{0,
  \log\frac{N n_{jk}}{n_j n_k}\right\}.
\end{equation}

We fit KL-divergence NMF at candidate ranks $K\in\{8,16,32\}$,

\begin{equation}
  X\approx WH, \qquad W,H\geq0,
\end{equation}

and symmetrize normalized left and right code loadings to obtain an $L_2$-normalized
procedure vector $\vect v_j\in\mathbb R_+^K$. We select rank by held-out procedure
retrieval and seed stability, not reconstruction error alone.

The provider's semantic state is a normalized revenue-weighted centroid over represented
procedures:

\begin{equation}
  \widetilde{\vect u}_{it}=\sum_j p_{ijt}\vect v_j,
  \qquad
  \vect u_{it}=\frac{\widetilde{\vect u}_{it}}
  {\lVert\widetilde{\vect u}_{it}\rVert_2}.
  \label{eq:state}
\end{equation}

We store dictionary dollar coverage and code count separately. This projection is not
an unconstrained quarterly factor fit and does not imply an expected dollar share.

\subsection{Temporal change}

The initial hypothesis considered a provider-specific transition
$\vect u_{i,t+1}=A_i\vect u_{it}+\vect\eta_{it}$. With eight quarterly observations, a
free $K\times K$ matrix is not identifiable. We therefore tested unchanged-state
persistence, a pooled diagonal transition with an entry vector, and a shrunken
provider-diagonal model:

\begin{equation}
  \widehat{\vect u}_{i,t+1}\propto D_i\vect u_{it}+\vect b_0.
\end{equation}

Because persistence won held-out comparisons, the promoted temporal measures are the
observed state cosine, $L_2$ movement, factor-wise contributions, family total variation,
and residual relative to persistence. We do not estimate or store $A_i$ as an expected
provider transition.

\subsection{First-use code profiles}

We evaluate a first-use event only for a provider observed before the current period.
We identify newly observed providers separately. For established providers, a code
event profile may include pre-adoption semantic compatibility

\begin{equation}
  c_{ijt}=\vect u_{i,t-1}^{\top}\vect v_j,
\end{equation}

peer prevalence, closest prior procedure, initial observed amount and share, companion-code
coherence, explicit predecessor use, and later persistence. The event taxonomy separates
provider-new, client-new but Medicare-established, simultaneous client/Medicare entry,
authoritative replacement, globally recent code, and metadata-unavailable events. We do
not promote a single plausibility score.

\subsection{Client and Medicare views}

The same fixed procedure dictionary and family mapping describe both observation systems:

\begin{equation}
  \vect u^{M}_{it}=\operatorname{project}(p^M_{it};V),\qquad
  \vect u^{C}_{it}=\operatorname{project}(p^C_{it};V).
\end{equation}

Cross-view outputs include family total variation, family cosine, semantic cosine,
dictionary coverage, client paid amount, Medicare reported amount, and baseline quarter.
Medicare context uses the latest completed quarter ending no later than the client window.
Cross-view distance describes selection or concentration, not provider anomaly.

\subsection{Final layered representation}

The promoted core is

\begin{equation}
  \mathcal D^{\mathrm{core}}_{it}=
  \left(L_{it},R_{it},P_{it},\Delta P_{it},C_{it}\right),
\end{equation}

where $L$ is scale, $R$ is transparent growth history, $P$ is the lineage-aware family
composition, $\Delta P$ is attributed change, and $C$ is support, billing, lineage, and
cross-view context. The optional semantic service adds $\vect u_{it}$ and procedure
vectors $\vect v_j$ for retrieval, fixed-width storage, and domains without a maintained
hierarchy.

\subsection{Domain ranking for review}

The same representation produces two queues without changing the evidence contract. A
national queue applies scale, mix, novelty, and semantic-frontier domains within Falcon's
proprietary Medicare panel. A client-specific queue applies those four domains to the
client view and adds client/Medicare mix divergence and growth disagreement. Scale
combines the percentile of absolute model-relative scale innovation with the percentile
of its attributed amount difference. Mix combines family total variation with mix-shift
amount, and novelty combines first-use amount share with first-use amount after excluding
left-censored histories and explicit code replacements. Semantic frontier combines the
unrepresented share with the amount not represented by the fixed dictionary. The two
additional client domains pair client/Medicare mix or growth disagreement with its
attributed amount. Percentiles are calculated
within sufficiently supported specialty groups, with a pooled fallback for small groups.
For unexpectedness percentile $U_d$ and materiality percentile $M_d$, domain $d$ is

\begin{equation}
  S_d=\sqrt{U_dM_d}.
\end{equation}

This construction requires both relative unusualness and material dollars. We retain
domain ranks as the primary review contract. The national queue uses

\begin{equation}
  S^{\mathrm{national}}=\max_{d\in\{\mathrm{scale,mix,novelty,frontier}\}} S_d,
\end{equation}

and orders providers by this value; attributed domain amount and current amount are
deterministic tie-breakers, not additional score components. For client queue ordering,
let $P$ be the largest domain score across the client and contextual views and $C$ the
strongest score from the other view. The bounded corroboration score is

\begin{equation}
  S^{\mathrm{review}}=P+0.25(1-P)C.
\end{equation}

Thus an independent contextual signal can support, but cannot overwhelm, the strongest
domain. The current national production contract requires the current quarter and its
exact immediate predecessor, at least four reported quarters, standard or high profile
support, at least \$10,000 of current allowed amount, and at least two observed service
families. Specialty groups with at least 30 eligible providers calibrate independently;
smaller groups share a pooled fallback. These thresholds and the amount contract are also
versioned with each build. National Medicare profiles use allowed amount, whereas client
profiles use paid-to-provider amount. The score is not a probability, predictive
accuracy measure, or allegation. Every ranked row retains amount, model and behavior
classes, attributed families/codes, support, coverage, and Medicare timing.

\begin{figure}[p]
  \centering
  \includegraphics[width=\linewidth,height=0.88\textheight,keepaspectratio]{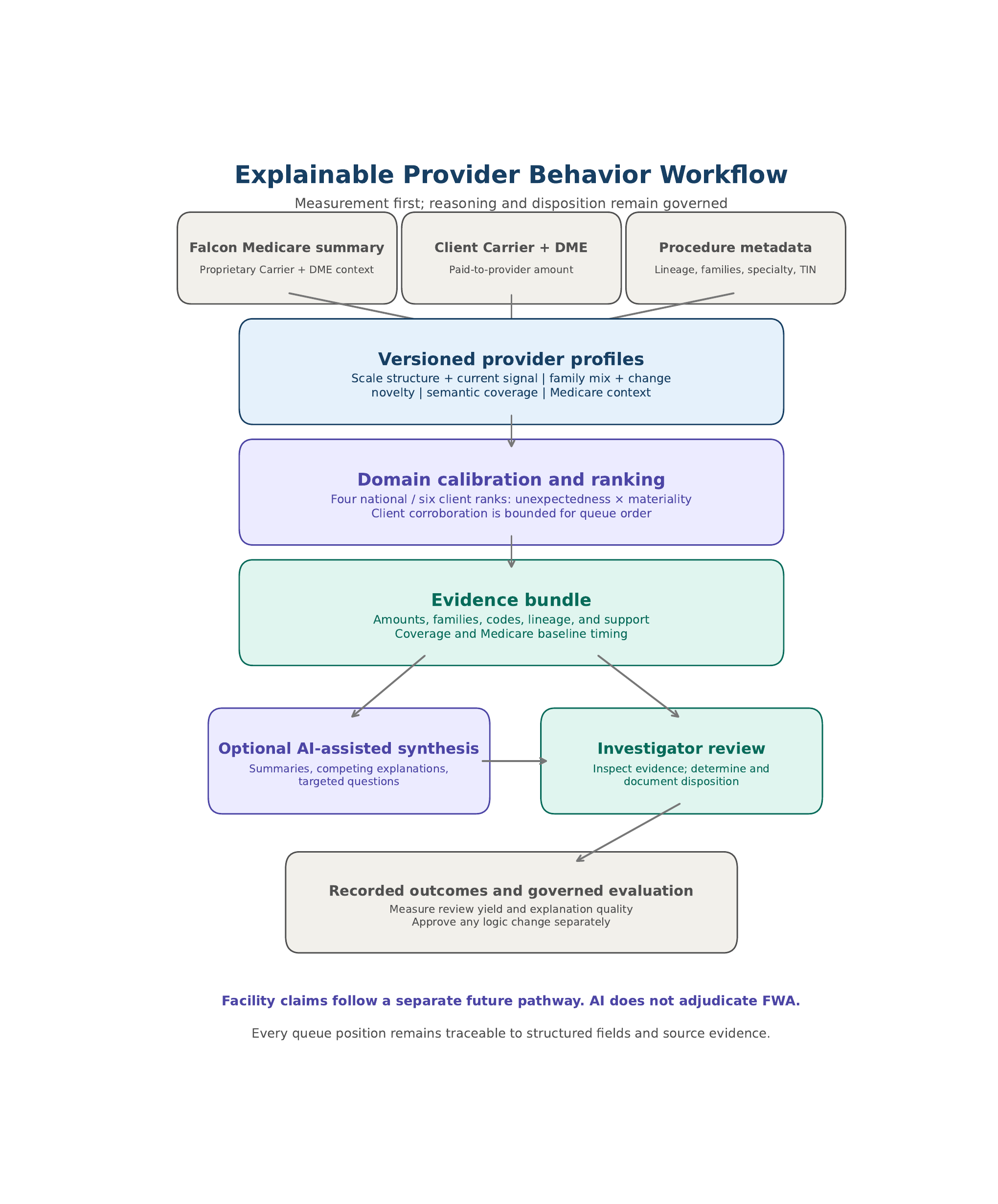}
  \caption{Operational workflow for theory-grounded provider review. Governed
  Carrier+DME inputs produce versioned behavior profiles, separately inspectable domain
  ranks, and an evidence bundle. AI synthesis is optional and remains downstream of
  measurement; investigator disposition and governance remain explicit.}
  \label{fig:workflow}
\end{figure}

Figure~\ref{fig:workflow} places these outputs in operational sequence. It shows that
measurement and ranking produce evidence for investigator review rather than a substitute
for it; optional AI synthesis remains downstream of the governed profile.

\section{Empirical Validation}

\subsection{Data support and sparsity}

Of 1,266,479 providers in the current Carrier+DME descriptor build, 590,885 (46.7\%)
appear in all ten quarters. The governed output contains 9,212,786 provider-quarter
profiles, while 253,002 providers (20.0\%) have fewer than four reported quarters and
therefore receive summary-only rather than individualized dynamics. These results support
sparse methods and explicit history-support fields.

\subsection{Scale descriptors}

The support-aware descriptive assignment is 38.2\% stable level, 19.3\% persistence,
19.8\% regularized AR(1), and 2.7\% drift; 20.0\% receive summary-only descriptors.
These labels summarize observed histories and do not forecast misconduct. The earlier
rolling scale experiments remain development evidence for choosing a simple temporal
summary, not a prospective validation target for the provider-profile algorithm.

\subsection{Semantic dictionary}

Direct masked NMF of provider revenue shares produced rank-32 held-out RMSE of 0.228,
compared with 0.154 for a code-mean baseline, and median cross-seed procedure-vector
cosine of 0.168. These results indicate that positive co-occurrence is more informative
for stable procedure geometry than reconstruction of the sparse share matrix.

\begin{figure}[ht]
\centering
\includegraphics[width=\linewidth]{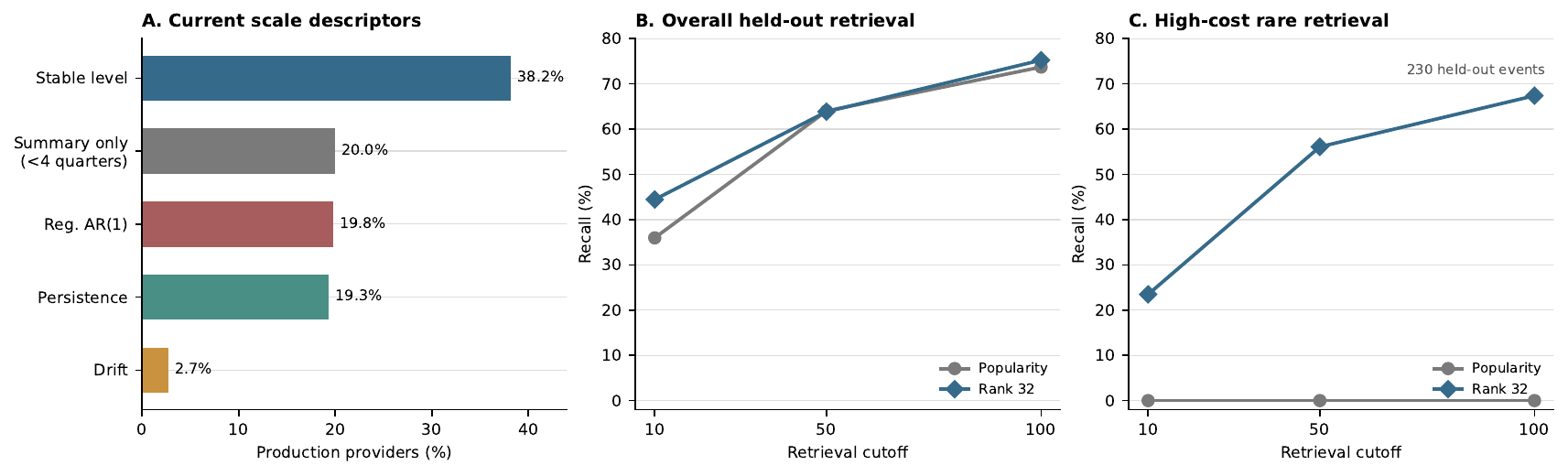}
\caption{Current 2026 Q2 production evidence supporting the layered representation.
Panel A shows support-aware scale-descriptor assignment across 1,266,479 Carrier+DME
providers. Panel B reports overall held-out procedure retrieval. Panel C isolates 230
high-cost rare held-out events, the primary semantic-dictionary objective.}
\label{fig:validation-overview}
\end{figure}

The selected dictionary contains 3,641 represented codes and 868,732 positive PPMI
associations; 50 additional supported codes have zero usable vectors and remain explicitly
uncovered. Pairwise geometry correlation across seeds is 0.951 and median top-10 neighbor
overlap is 60\%. The selected rank-32 model improves overall recall at 10 from 36.0\%
under popularity ranking to 44.4\%. Its declared objective gives greater weight to
high-cost rare codes: on 230 such held-out events, popularity has zero recall through 100
candidates, while the selected model reaches 23.5\% at 10, 56.1\% at 50, and 67.4\% at
100. Overall recall at 50 is nearly equal to popularity (63.8\% versus 64.0\%), so the
rare-code gain remains an objective tradeoff rather than universal dominance. The full
retrieval comparison appears in Figure~\ref{fig:validation-overview}.

\subsection{Historical alternative-model audits}

Earlier development experiments compared persistence with pooled and provider-specific
semantic transition operators and compared composite first-use plausibility scores with
their individual evidence fields. Neither additional model supplied stable incremental
explanatory value. We therefore retain observed movement, lineage, prevalence, coverage,
amount, and support as separately inspectable descriptors. These experiments motivated
simplification; they are not presented as prospective prediction by the production
algorithm.

\subsection{Quarterly refresh stability}

R016 rebuilt the label-independent review lanes for 2025 Q4, 2026 Q1, and 2026 Q2 using
only history support, model selection, first-use events, and specialty calibration available
through each quarter. It exactly reproduced the 414,093-provider current production
population and its scores before comparing earlier landmarks. The eligible population
retained 81.8\% of its providers from Q4 to Q1 and 90.8\% from Q1 to Q2. Adjacent-quarter
review-rank correlations were 0.262 and 0.237, while exact mix-class agreement was 57.1\%
and 56.3\%. The top-1,000 Jaccard overlap was 21.1\% and 22.6\%; the broader top-1\%
overlap was 13.5\% and 14.0\%. The eligibility population is moderately persistent, but
the ranked tail is deliberately sensitive to quarter-specific materiality, first-use events,
and peer calibration. These are stability statistics, not predictive accuracy.

\begin{figure}[ht]
\centering
\includegraphics[width=\linewidth]{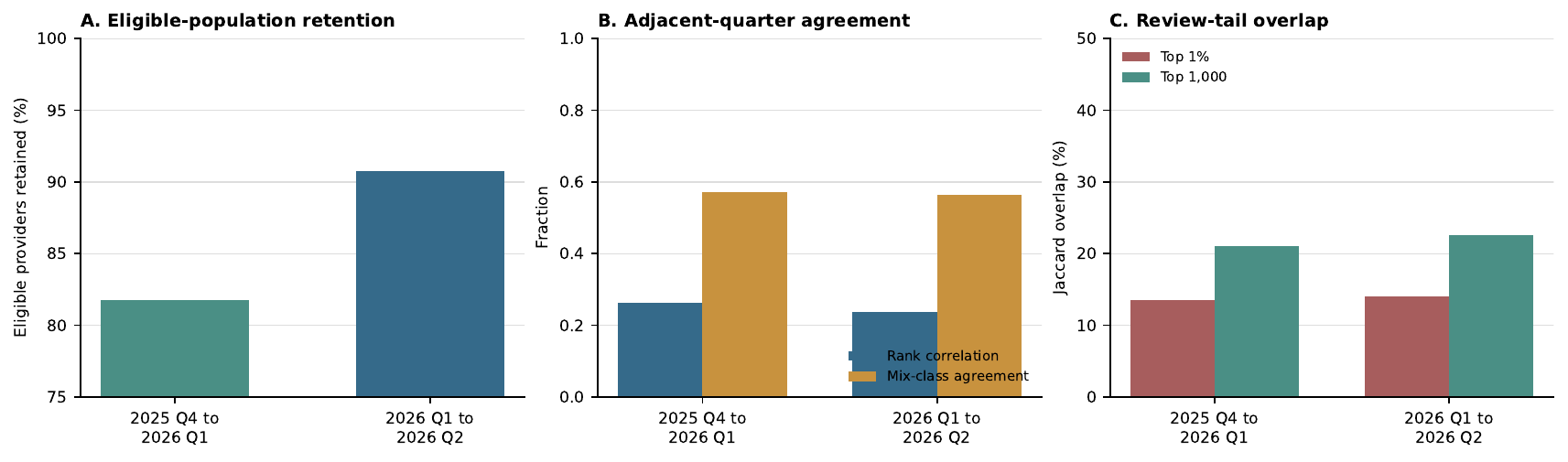}
\caption{Descriptive refresh stability across the two newly evaluated quarter transitions.
Panel A reports retention of the prior eligible population. Panel B reports review-rank
correlation and exact mix-class agreement among overlapping providers. Panel C reports
Jaccard overlap in the review tail. Movement can reflect changed provider behavior,
population composition, or procedure coverage; it is not an outcome forecast.}
\label{fig:refresh-stability}
\end{figure}

\subsection{Eight-client completed-quarter replication}

All eight managed client panels were rebuilt at the same completed 2026 Q2 cutoff in
isolated research schemas. Together they contain 894,849 provider profiles; providers may
appear in more than one panel, so this is not a unique-provider count. Amount-weighted
semantic coverage ranges from 86.9\% to 95.2\%. Review eligibility ranges from 0.6\% to
4.9\%, reflecting different observed populations rather than an expected common rate.

\begin{table}[ht]
\centering
\caption{Common completed-2026-Q2 client replication.}
\label{tab:client-replication}
\begin{tabular}{lrrr}
\toprule
Panel & Provider profiles & Review eligible & Semantic coverage \\
\midrule
A & 8,976 & 151 & 92.2\% \\
B & 12,324 & 245 & 86.9\% \\
C & 49,887 & 459 & 90.8\% \\
D & 28,847 & 165 & 95.2\% \\
E & 9,846 & 72 & 94.9\% \\
F & 55,065 & 892 & 95.0\% \\
G & 176,651 & 3,878 & 90.2\% \\
H & 553,253 & 27,021 & 89.5\% \\
\bottomrule
\end{tabular}
\end{table}

\subsection{Simplicity stress test}

The effective-dated concept layer maps 99.78\% of Medicare dollars. A family profile
needs two items at the median and four at the 90th percentile to explain 90\% of a
Medicare provider-quarter's dollars. Across 5.77 million matched transitions, family
total-variation movement correlates 0.790 with PPMI movement. Under removal of the
smallest cells totaling up to 10\% of dollars, the 10th-percentile within-quarter family
cosine remains 0.9951, compared with 0.9982 for PPMI. The learned representation is
marginally smoother, but the simple family representation is more direct for behavioral
explanation.

\begin{table}[H]
\centering
\caption{Layered representation decision.}
\label{tab:layers}
\begin{tabularx}{\linewidth}{@{}>{\raggedright\arraybackslash}p{0.19\linewidth}YY@{}}
\toprule
Layer & Strength & Promoted role \\
\midrule
Literal codes & Exact billing evidence; maximal specificity & Drill-down and attribution \\
Lineage/family shares & Compact, stable, interpretable; handles replacements & Core provider profile \\
Rank-32 PPMI state & Fixed width; semantic retrieval; smoother sparse vocabulary & Optional semantic service \\
AR/transition models & Diagnostic temporal summaries & Not required in core \\
\bottomrule
\end{tabularx}
\end{table}

\subsection{Frozen retrospective administrative association check}
\label{sec:external-validation}

As a secondary retrospective construct check, we examined whether label-independent
descriptors were associated with later recorded administrative actions in a previously
frozen rolling-origin comparison with the CMS revoked-provider and HHS OIG LEIE endpoints described in
Section~\ref{sec:external-endpoints}. These lists were
used only for post-hoc evaluation. No list membership, reason, effective date, or derived
label entered the procedure dictionary, provider representation, score construction, or
queue ranking.

For each of five landmarks, a rank-32 PPMI dictionary was relearned from only the four
preceding PSPS service quarters. Current and prior provider states were projected through
that past-only geometry, and every landmark ended at least 180 days before the
administrative effective date. The evaluation therefore clamps PSPS service chronology
without using either endpoint to fit a representation or construct a score.

The resulting cohorts contain 83 revoked-provider and 56 LEIE outcomes. Absolute scale
change produced population AUROC 0.632 and 0.686 respectively, with matched concordance
0.648 [0.576, 0.716] and 0.698 [0.611, 0.777]. Literal and rolling semantic movement were
weaker. For revoked-provider retrieval, the materiality-adjusted top 250 per quarter found
six outcomes in 1,000 reviews (0.600\% endpoint precision, 7.23\% recall, and 218-fold
lift over random). None of those six appeared in the current LEIE snapshot. LEIE had no
hits within the top 1,000 providers per quarter; the scale top 5\% found 16 outcomes in
191,431 rows (0.00836\% precision, 28.57\% recall, and 5.71-fold lift), including five
not present in the current revoked snapshot.

A regulatory-code audit shows substantial endpoint heterogeneity. The 83 revoked cases
comprise 27 standards or enrollment noncompliance records, 17 on-site review or reporting
records, 10 undue-risk affiliation records, 10 felony records, nine licensure or
prescribing records, eight exclusion or program-termination records, and two billing-abuse
or false-information records. Five of the six top-250 hits are from the on-site/reporting
family and the sixth is a billing-abuse pattern-or-practice action. The extreme revoked
lift therefore describes a narrow administrative mechanism, not uniform performance over
revocation reasons. LEIE scale retrieval is more heterogeneous: its 16 top-5\% hits
include eight license actions, four program-related crimes, and four healthcare-fraud or
false-statement exclusions.

Because these endpoints are rare, observed lift is intrinsically discrete. Ten thousand
case-resampling replicates, stratified by landmark quarter and holding review capacity
fixed, give a 95\% interval of 72.8--400.5-fold for revoked top-250 lift and
3.57--8.21-fold for LEIE top-5\% lift. Leave-one-landmark estimates range from
121.7--303.4-fold and 5.45--5.93-fold respectively. These intervals condition on the
observed administrative cohorts; they do not include endpoint-ascertainment error,
unobserved fraud, or reason-code error.

A label-independent score audit also identifies substantial ties. Raw semantic-frontier
share is zero in approximately 96\% of risk-set rows. Its \texttt{CUME\_DIST} transform
places that minimum-value tie near the 96th percentile, and the behavioral maximum then
has a single tied group containing approximately 89\% of rows. Amount and provider
identifier determine the declared ordering within such ties. We preserve the frozen R015
scores here and report this limitation rather than retrofitting the method after examining
the endpoints; tie-safe calibration is a future, separately frozen algorithmic change.

\begin{figure}[p]
  \centering
  \includegraphics[width=\linewidth,height=0.84\textheight,keepaspectratio]{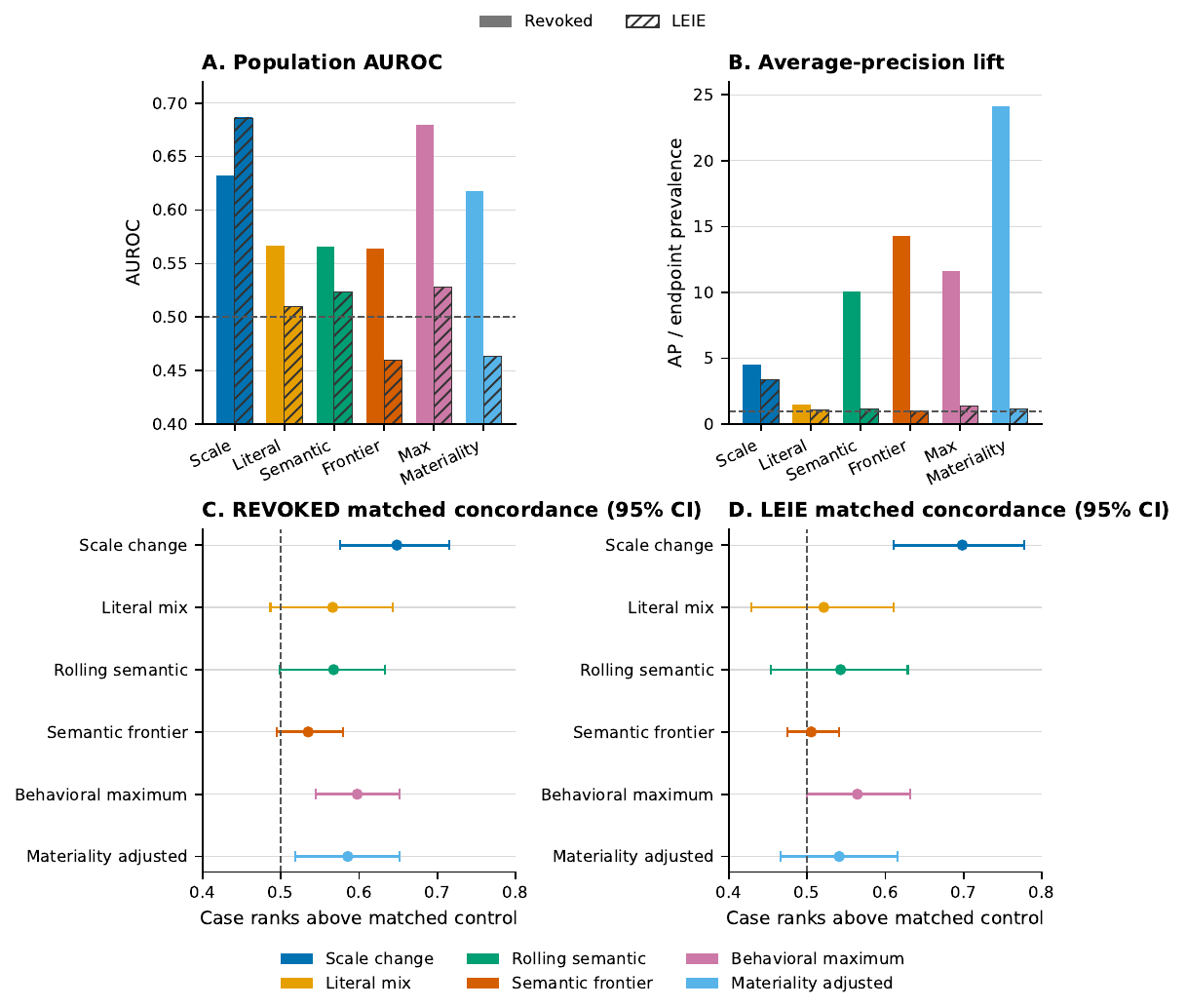}
  \caption{Rolling-origin external discrimination for the revoked-provider and LEIE
  endpoints. Colors identify the six label-independent score lanes; solid and hatched
  bars distinguish endpoints. Dashed reference lines mark chance AUROC or concordance
  (0.5) and random average-precision lift (1.0). Matched intervals are case-bootstrap
  95\% intervals.}
  \label{fig:external-discrimination}
\end{figure}

\begin{figure}[p]
  \centering
  \includegraphics[width=\linewidth,height=0.84\textheight,keepaspectratio]{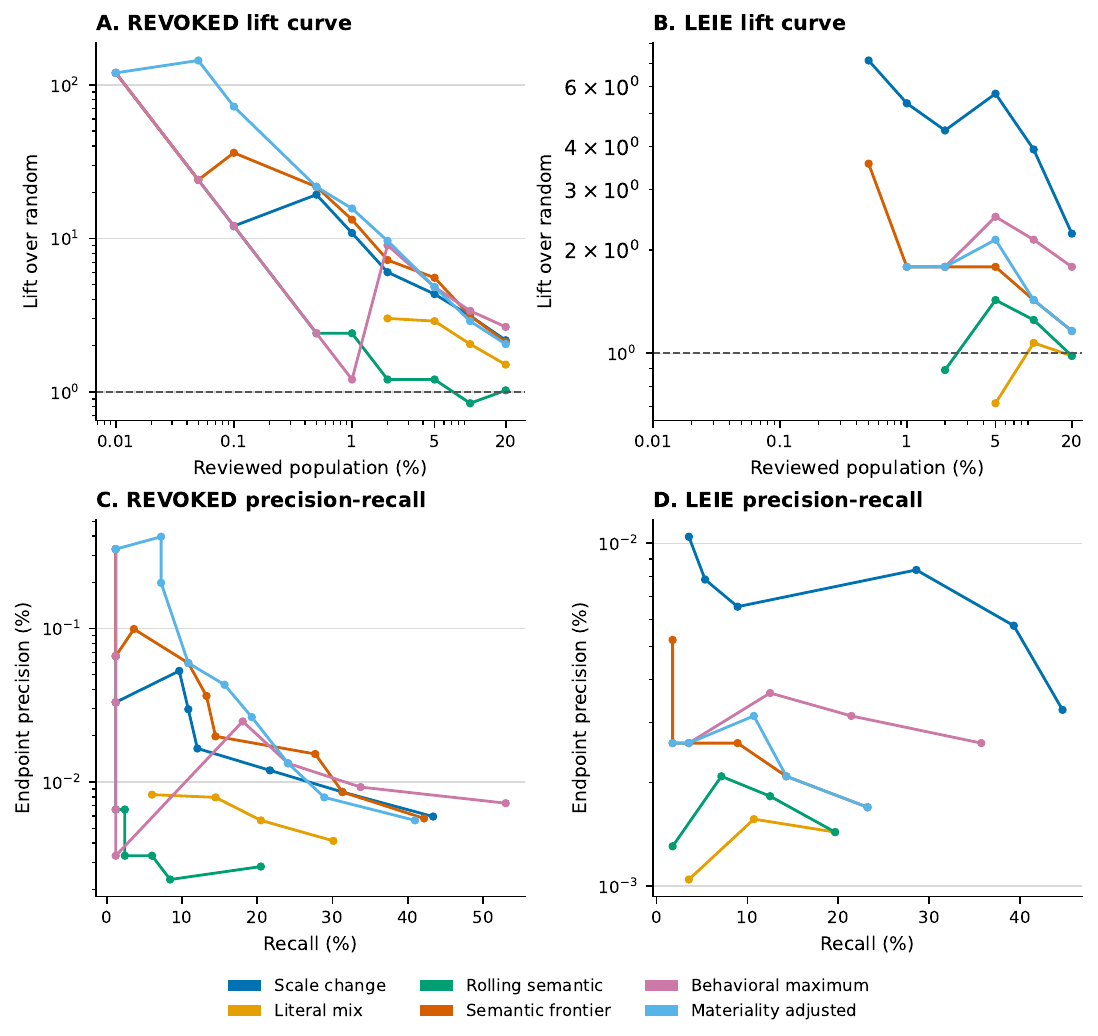}
  \caption{Endpoint retrieval across review capacity. Top panels show lift relative to
  within-quarter random selection; bottom panels show endpoint precision against recall.
  Markers are observed capacities and connecting segments are visual guides, not fitted
  smoothers; absent extreme-tail markers have zero retrieved outcomes.
  Axes use logarithmic scaling where needed because the administrative endpoints are
  rare. These are positive-unlabeled administrative outcomes, so precision is endpoint
  precision rather than estimated true-fraud precision.}
  \label{fig:external-retrieval}
\end{figure}

\begin{figure}[p]
  \centering
  \includegraphics[width=\linewidth,height=0.84\textheight,keepaspectratio]{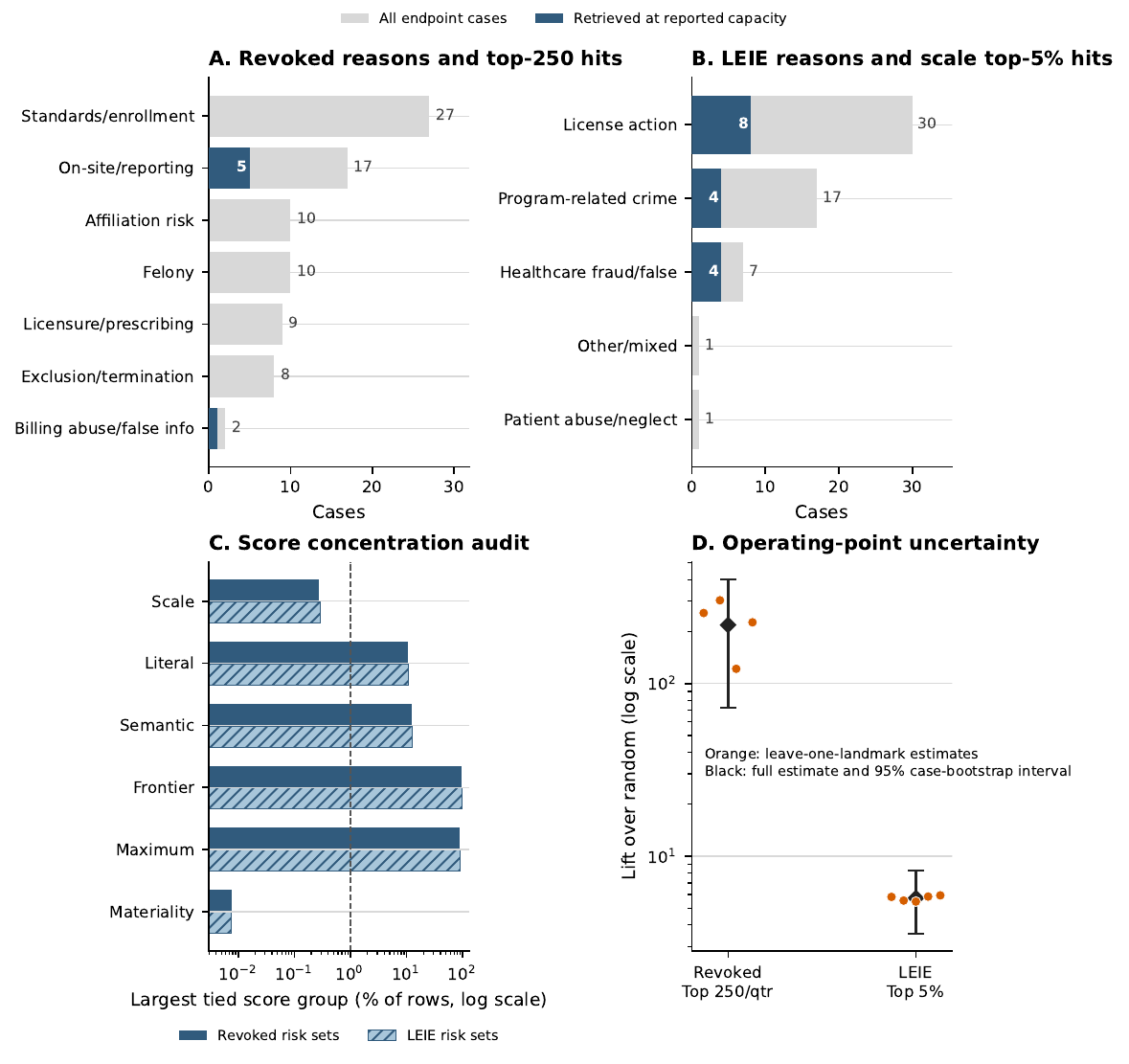}
  \caption{Evaluation audit without score refitting. Panels A and B compare the full
  reason composition with outcomes retrieved at the two reported operating points.
  Panel C reports the mean largest tied score group across landmark risk sets; the dashed
  line marks 1\%. Panel D shows the full lift estimate, a 95\% landmark-stratified
  case-bootstrap interval, and each leave-one-landmark estimate. The logarithmic scale is
  necessary because the endpoint prevalences and review capacities differ sharply.}
  \label{fig:external-audit}
\end{figure}

Figure~\ref{fig:external-discrimination} tests whether association is consistent across
score lanes and discrimination metrics. Scale change is the most consistent discriminator
across the two endpoints; large average-precision lift ratios must still be read against
very low endpoint prevalence rather than as high absolute precision.
Figure~\ref{fig:external-retrieval} translates discrimination into review capacity. The
revoked-provider lift is strongest but discrete at narrow capacities, whereas LEIE
retrieval becomes meaningful over a broader review fraction; missing extreme-tail points
indicate zero retrieved outcomes, not omitted experiments.
Figure~\ref{fig:external-audit} asks how far those operating-point results generalize. It
shows concentration in particular revocation reasons, substantial score ties, and wide
revoked-provider uncertainty, all of which limit interpretation of the headline lift.

The lists observe only a process- and capacity-selected fraction of the underlying
misconduct universe and also include actions unrelated to proven fraud. They therefore
neither enumerate all actual fraud nor consist entirely of adjudicated fraud cases.
Absence from a list does not clear a provider, membership does not by itself establish
fraud, and list-based recall is not recall of all fraud. The appropriate interpretation is
objective association with later administrative outcomes and expansion of a supported
investigative queue, not adjudication or automated adverse action.

\section{Provider Case Studies}

Claims can identify change, but claims alone cannot establish intent. FWA review requires
investigation and supporting research. The system should therefore find material,
unusual, or unexplained behavior and preserve enough evidence for an analyst to ask the
next question. We examine two complementary queues. The national queue uses only Falcon's
proprietary Medicare summary and asks which providers are unusual within that broad
reference view. The client queue is restricted to providers observed by one payer and
adds within-client change and client/Medicare disagreement. The examples are anonymized,
selected to span distinct evidence domains, and are not a prevalence estimate or an
allegation. Each profile was generated by the same versioned rules used for its queue.

\subsection{Cross-view replication}

The common-quarter client builds preserve the client and proprietary-Medicare views as
separate measurements. Across the eight panels, exact mix-behavior disagreement among
comparable profiles ranges from 62.9\% to 76.2\%, and scale-behavior disagreement ranges
from 54.2\% to 74.6\%. Agreement is not the target: the two views cover different patient
populations, payer selection, and timing. A nationally stable mix may therefore appear as
gradual rebalancing or an emerging service line within one client. The disagreement is a
contextual question for review, not an error rate or FWA determination.

\begin{figure}[ht]
\centering
\includegraphics[width=\linewidth]{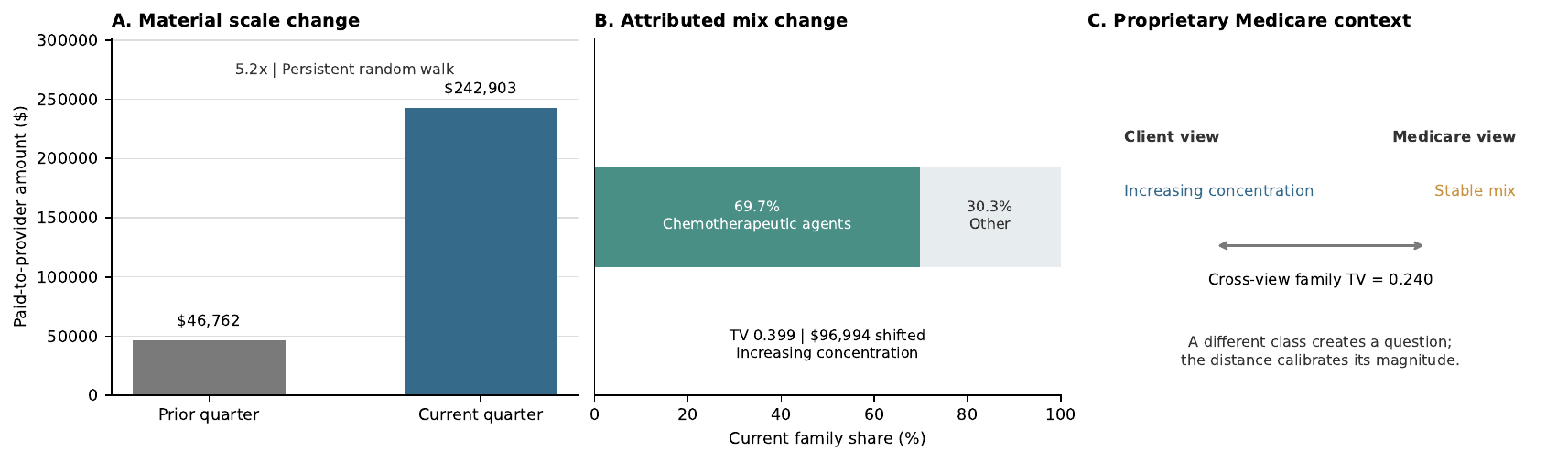}
\caption{Anatomy of an anonymized client provider profile. Scale combines growth with
current amount; mix combines family movement with attributed dollars; and the
client/Medicare comparison pairs readable classes with a continuous distance. These
fields formulate review questions and do not constitute an FWA verdict.}
\label{fig:profile-anatomy}
\end{figure}

Figure~\ref{fig:profile-anatomy} illustrates how the fields are read together for one
anonymized supported profile. The 5.2-fold increase is not ranked in isolation: current
amount establishes materiality, attributed family shares explain the concentration, and
the proprietary Medicare view supplies a separate contextual comparison. The class
difference is descriptive; the continuous cross-view distance calibrates its magnitude.

\subsection{National proprietary Medicare-summary cases}

The national examples use the latest complete quarter in the refreshed validation build,
2026 Q2, with Carrier and DME summarized at rendering-NPI by HCPCS grain. Their selection
draws only on this national PSPS view, not on client-specific claims.

\begin{caseexample}{Scale and service-line change converge}
A nurse-practitioner profile increased in reported Medicare amount from \$6,256 to
\$1.08 million. Family total variation was 1.000 and \$1.06 million was attributed to a
newly observed skin-allograft family, which represented 98.2\% of current amount; skin
grafting and debridement supplied the remaining material introduced families. Scale,
mix, amount, and specialty context therefore converge on a focused review question: what
clinical, organizational, coding, or billing change explains this new service line? The
profile identifies the question; it does not answer it.
\end{caseexample}

The following two lineage examples are retained from the earlier 2025 Q4 qualitative
audit because their purpose is to illustrate representation behavior rather than report
current queue prevalence or performance.

\begin{caseexample}{Code change without clinical change}
A nurse-practitioner profile increased from \$26.00 million to \$93.38 million. The
dominant code changed from Q4275 to Q4344, but both map to the skin-allograft family.
Consequently, code-level novelty was large while family total variation was only 0.001
and the mix class remained stable. The scale increase remains material and reviewable,
but the family representation prevents the code change alone from being misread as entry
into an unrelated clinical domain. This is precisely why literal codes and clinical
families are retained together. A second profile moved from G2012 to 98016 with similar
amount and stable scale. Here the structured CMS crosswalk established a direct
replacement even though the semantic vectors were not close. Together, the examples
show why clinical family and authoritative lineage take priority over an embedding when
they provide stronger evidence.
\end{caseexample}

\begin{caseexample}{Stable behavior at the semantic frontier}
A preventive-medicine profile decreased modestly from \$15.32 million to \$14.40 million
and had family total variation of 0.0003. Yet semantic dollar coverage was only 0.001:
Q4279, representing \$14.38 million and 99.9\% of current amount, was outside the learned
procedure dictionary. This case ranks in the semantic-frontier domain even though scale
and family mix are currently stable. Low semantic coverage is therefore evidence about
the limits of the model vocabulary---especially for rare or new high-cost products---not
evidence that the provider changed or behaved improperly.
\end{caseexample}

\subsection{Client-specific cases}

The Client B examples use a complete 2026 Q2 quarter (April 1--June 30), restricted to
active Carrier and DME claims and nonnegative paid-to-provider amount. The refreshed build
produced 12,324 provider profiles, of which 245 met the declared review-support rule.

\begin{caseexample}{Stable client behavior, different Medicare view}
A provider recorded \$95,070 in Q2 after \$135,092 in Q1. Its within-client family mix
was nearly unchanged (total variation 0.001), yet the client and Medicare family profiles
were almost disjoint (total variation 0.999). The primary domain was therefore
cross-view mix rather than client mix. This is the intended question-generating use of
the proprietary Medicare reference: why does the client's observed patient spend
concentrate in services that differ so strongly from the provider's broader Medicare
activity? Network selection, patient mix, contracting, and timing remain plausible
explanations.
\end{caseexample}

\begin{caseexample}{Client-new does not necessarily mean provider-new}
A radiation-oncology provider showed a material service newly observed in the client's
window, but the same service was established in the provider's proprietary Medicare
history. The profile therefore distinguishes payer-panel
entry from provider-level service adoption. That distinction changes the follow-up from
``why did this provider begin the service?'' to ``why did this service newly appear for
this client population?''
\end{caseexample}

Together, the cases show the intended role of the system: domain ranks
surface a reason for review; structural and current labels explain the pattern; attributed
families, codes, and amounts show what changed; and the proprietary Medicare reference
supplies contextual evidence. None of these fields infers intent.

\section{Discussion}

\subsection{Why description precedes detection}

The main result is architectural rather than algorithmic. Provider understanding does
not require a universally optimal prediction model. It requires an explicit separation
of scale, composition, temporal change, and observation context. A downstream queue can
prioritize domain-specific combinations of unexpectedness and material amount: scale
innovation, family movement, first use, client/Medicare mix or growth disagreement, and
semantic coverage. Because the underlying components remain visible, analysts can
distinguish a small-base ratio from a large-dollar expansion, a literal code change from
an authoritative replacement, and provider-wide change from client selection.

This architecture complements rather than replaces Falcon's other FWA methods. Its role
is to produce a transparent provider descriptor and an interpretable review funnel that
can be combined with other rules, models, clinical knowledge, network evidence, and
investigative workflows. A provider absent from this queue is not thereby cleared, and a
provider ranked highly is not thereby adjudicated.

This layered representation also avoids a common failure mode of anomaly systems: a
single score can rank a case without revealing whether the case is actionable. Our queue
therefore publishes domain scores and ranks first. The national queue orders the maximum
of its four domain scores. The client queue's combined score is only a bounded ordering
rule in which the strongest domain dominates and an independent second view adds limited
corroboration. Both derive from versioned, support-qualified descriptors and must be
validated against a declared workflow.

Published payment-integrity vendors describe similar ingredients---rules, network
analysis, and predictive or generative AI scoring---but, to our knowledge, none has
released a technical account with an explicit measurement contract, formal decomposition,
and held-out validation statistics. We view that as a gap the broader payment-integrity
community can help close, and it is the reason we publish this component on its own.

The post-hoc administrative validation in Section~\ref{sec:external-validation}
reinforces this boundary. The descriptors are associated with later revocation and
exclusion outcomes. Neither list contributes to representation learning, score
construction, or ranking. At the same time, limited list coverage, mixed administrative
reasons, the wide revoked-provider interval, and the score-tie audit show why the result
cannot be interpreted as fraud classification. The evidence supports a broader
investigative funnel; it does not convert the profile into a verdict.

\subsection{Limitations}

The proprietary Medicare summary may omit low-volume cells under its construction and
does not establish true zeros. Ten quarters remain insufficient for rich provider-specific dynamics. Revenue shares do not
adjust for patient case mix, beneficiary exposure, geography, contracting, or clinical
necessity. Procedure hierarchies are incomplete and change over time. Enforcement labels
are used only for post-hoc external validation, never representation fitting or score
construction; the results therefore establish administrative-outcome association, not
fraud-discrimination performance. Case studies are deliberately selected
from a supported review queue and cannot establish prevalence or review yield. Client findings depend on payer-specific data
contracts, and we must revalidate them after any revenue-field or claims-status change.

\subsection{Operational safeguards}

Production outputs record source table, included claim types, accepted claim status, amount column, cutoff,
dictionary version, build identifier, and validation results. We exclude future-dated
service rows and expose explicit completion fractions for partial client quarters. We evaluate first-use
codes only for providers observed before the current window. Cross-view comparisons require a
completed Medicare baseline and explicit coverage. These safeguards are part of the
method, not implementation details.

Client data is also treated independently by design. The semantic dictionary, lineage
rules, and scale model are fit once on Falcon's proprietary PSPS panel and are never
retrained or refit on client claims; a client engagement only applies these fixed
components to score and explain its own data, and nothing learned from that data feeds
back into them. One client's data therefore never becomes a training input for another
client's analysis.

\subsection{AI-assisted reasoning after measurement}

AI presents a new opportunity once it is given a governed, well-structured description
of provider behavior. Rather than asking a model to infer intent from raw claims or from
a single anomaly score, the intelligence layer can synthesize explicit scale, mix,
lineage, semantic-coverage, and Medicare-comparison evidence. It can retrieve relevant
codes and policies, articulate competing explanations, identify missing context, and
propose targeted questions for an investigator. Each statement must remain traceable to
the versioned profile and its source evidence.

This role does not convert descriptive signals into a fraud verdict. Claims do not reveal
clinical necessity or intent, and generated explanations can be incomplete or wrong.
Investigators therefore review the evidence directly, record dispositions separately,
and use governed evaluation to improve mappings, thresholds, and retrieval. The AI layer
assists evidence synthesis; it does not define provider behavior, adjudicate FWA, or
replace human review. Its position downstream of governed measurement is shown in
Figure~\ref{fig:workflow}.

\section{Future Directions}

The immediate next step is a blinded multi-analyst comparison of literal, family, and
semantic profiles. Reviewers should score explanation quality, time to explanation,
missing context, and whether the review requires another data pull. This is a more relevant
endpoint than reconstruction loss alone.

Several extensions follow naturally:

\begin{enumerate}[leftmargin=*]
  \item \textbf{Investigator workflow evaluation.} Compare review yield, explanation
        quality, investigator agreement, time to explanation, and additional-data burden
        with existing rules, without treating the profile as an outcome predictor.
  \item \textbf{Improved peer context.} Construct peers using specialty, service family,
        scale, geography, and organizational relationships, while keeping peer distance
        separate from the provider's own temporal change.
  \item \textbf{Exposure-aware client comparison.} Incorporate member months, attribution,
        network participation, and benefit design to distinguish payer selection from
        provider behavior.
  \item \textbf{Additional domains.} Apply the same entity--time scale, item composition,
        attributed change, and optional semantic-geometry interface to diagnoses, drugs,
        sites of care, referral partners, facilities, and billing entities. A diagnosis
        prototype already transfers without architectural redesign.
  \item \textbf{Graph context.} Connect provider profiles to patient, billing-entity,
        referral, pharmacy, and facility networks. Graph anomaly measures can then enrich
        rather than replace the longitudinal profile.
  \item \textbf{Change-point and duration models.} Classify changes as transient,
        intermittent, or sustained after sufficient history accumulates; distinguish
        emerging-practice ramp from mature-provider acceleration.
  \item \textbf{Tie-safe calibration and persistent change.} Replace upper-ECDF handling
        of minimum-value ties with a declared tie-safe rank transform, recalibrate combined
        lanes without endpoint labels, and compare one-quarter movement with trailing-median
        or sustained multi-quarter change in a separately frozen run.
  \item \textbf{Uncertainty and fairness.} Quantify descriptor uncertainty under sparse
        data and evaluate whether support thresholds or peer definitions systematically
        change review rates across provider types or geographies.
\end{enumerate}

\section{Conclusion}

This study evaluates provider profiling as governed descriptive measurement, not as
prospective prediction. The refreshed ten-quarter national build and common-quarter
eight-client replication show that the representation can be recomputed reproducibly,
quantify both eligible-population persistence and ranked-tail sensitivity, and remain
explicit about coverage and population change. The operational questions are descriptive: how large is the provider,
what services define the provider, what changed, and what lineage or cross-view context
accompanies that change?

The resulting framework is intentionally simple at its core. Transparent scale history
and lineage-aware family shares explain ordinary provider behavior and major transitions.
A learned PPMI geometry adds genuine value for semantic retrieval and fixed-width machine
representation, but it is optional rather than mandatory. This separation makes the
system extensible without making every analyst explanation depend on a latent model. The
profile is therefore best understood as evidence infrastructure for FWA review: a
reproducible way to characterize providers and surface interpretable changes, not a claim
that statistical unusualness establishes fraud, waste, abuse, or intent.
It is one engine within Falcon's broader, multi-angle identification and investigation
system, designed to combine with and extend into the rules, network, and statistical
engines around it, rather than a proposed universal FWA detector.

\begingroup
\small
\sloppy
\bibliographystyle{plainnat}
\bibliography{provider_behavior_references}
\endgroup

\end{document}